\documentclass[aps,prd,twocolumn,groupedaddress,amsmath,amssymb]{revtex4-2}

\usepackage{amssymb}
\usepackage{overpic}
\usepackage{graphicx}
\usepackage{subfigure}
\usepackage{dcolumn}% Align table columns on a decimal point
\usepackage{bm}% bold math
\usepackage{epsfig}
\usepackage{xcolor}
\usepackage{enumerate}
\usepackage{booktabs}
\usepackage{multirow}
\usepackage{hyperref}
\hypersetup{hidelinks,
	colorlinks=true,
	allcolors=blue,
	pdfstartview=Fit,
	breaklinks=true}

\begin{document}

\newcommand{\jpsi}{J/\psi}
\newcommand{\pip}{\pi^+}
\newcommand{\pin}{\pi^-}
\newcommand{\pio}{\pi^0}
\newcommand{\g}{\gamma}
\newcommand{\gev}{GeV/c$^2$}
\newcommand{\mev}{MeV/c$^2$}
\newcommand{\ar}{\rightarrow}
\newcommand{\ks}{K_S^{0}}
\newcommand{\etap}{\eta^\prime}
%\let\oldequation\equation
%\let\oldendequation\endequation
%\renewenvironment{equation}{\linenomathNonumbers\oldequation}{\oldendequation\endlinenomath}

%Title of paper
\title{\Large \boldmath \bf Improved measurements of the Dalitz decays $\eta/\eta'\rightarrow\gamma e^{+}e^{-}$}

\author{
\begin{small}
  \begin{center}
%% Saved at => 2023-09-11
M.~Ablikim$^{1}$, M.~N.~Achasov$^{4,b}$, P.~Adlarson$^{75}$, O.~Afedulidis$^{3}$, X.~C.~Ai$^{80}$, R.~Aliberti$^{35}$, A.~Amoroso$^{74A,74C}$, Q.~An$^{71,58}$, Y.~Bai$^{57}$, O.~Bakina$^{36}$
% ...
\\
\vspace{0.2cm}
(BESIII Collaboration)\\
\vspace{0.2cm} {\it
$^{1}$ Institute of High Energy Physics, Beijing 100049, People's Republic of China\\
% ...
$^{3}$ Bochum  Ruhr-University, D-44780 Bochum, Germany\\
$^{4}$ Budker Institute of Nuclear Physics SB RAS (BINP), Novosibirsk 630090, Russia\\
% ...
$^{35}$ Johannes Gutenberg University of Mainz, Johann-Joachim-Becher-Weg 45, D-55099 Mainz, Germany\\
$^{36}$ Joint Institute for Nuclear Research, 141980 Dubna, Moscow region, Russia\\
% ...
$^{57}$ Southeast University, Nanjing 211100, People's Republic of China\\
$^{58}$ State Key Laboratory of Particle Detection and Electronics, Beijing 100049, Hefei 230026, People's Republic of China\\
% ...
$^{71}$ University of Science and Technology of China, Hefei 230026, People's Republic of China\\
% ...
$^{74}$ University of Turin and INFN, (A)University of Turin, I-10125, Turin, Italy; (B)University of Eastern Piedmont, I-15121, Alessandria, Italy; (C)INFN, I-10125, Turin, Italy\\
$^{75}$ Uppsala University, Box 516, SE-75120 Uppsala, Sweden\\
% ...
$^{80}$ Zhengzhou University, Zhengzhou 450001, People's Republic of China\\

\vspace{0.2cm}
% ...
$^{b}$ Also at the Novosibirsk State University, Novosibirsk, 630090, Russia\\
% ...

}
%% ends here %%

\end{center}
\end{small}
}

%####################################     abstract
\begin{abstract}
Based on a data sample of 10 billion $J/\psi$ events collected with the BESIII detector, improved measurements of the Dalitz decays $\eta/\eta'\rightarrow\gamma e^+e^-$ are performed, where the $\eta$ and $\eta'$ are produced through the radiative decays $J/\psi\rightarrow\gamma \eta/\eta'$. The branching fractions of $\eta\rightarrow\gamma e^+e^-$ and $\eta'\rightarrow\gamma e^+e^-$ are measured to be $(7.07 \pm 0.05 \pm 0.23)\times10^{-3}$ and $(4.83\pm0.07\pm0.14)\times10^{-4}$, respectively. Within the single-pole model, the parameter of electromagnetic transition form factor for $\eta\rightarrow\gamma e^+e^-$ is determined to be $\Lambda_{\eta}=(0.749 \pm 0.027 \pm 0.007)~ {\rm GeV}/c^{2}$. Within the multipole model, we extract the electromagnetic transition form factors for $\eta'\rightarrow\gamma e^+e^-$ to be  $\Lambda_{\eta'} = (0.802 \pm 0.007\pm 0.008)~ {\rm GeV}/c^{2}$ and $\gamma_{\eta'} = (0.113\pm0.010\pm0.002)~ {\rm GeV}/c^{2}$. The results are consistent with both theoretical predictions and previous measurements. The characteristic sizes of the interaction regions for the $\eta$ and $\eta'$ are calculated to be $(0.645 \pm 0.023 \pm 0.007 )~ {\rm fm}$ and $(0.596 \pm 0.005 \pm 0.006)~ {\rm fm}$, respectively.  In addition, we search for the dark photon in $\eta/\eta^\prime\rightarrow\gamma e^{+}e^{-}$, and the upper limits of the branching fractions as a function of the dark photon are given at 90\% confidence level.
\end{abstract}

\maketitle

%####################################    introduction
\section{Introduction}
In the  Dalitz decays $\eta/\eta^\prime\to\gamma l^+l^- (l=e,\mu)$, the lepton pair is formed through the internal conversion of an intermediate virtual photon.  These decays are of special interest since their decay rates are sensitive to the electromagnetic structure arising at the vertex of the transition. 
Deviations of the measured quantities from their quantum electrodynamics (QED) predictions are usually described in terms of a timelike transition form factor, which sheds light on the meson's structure~\cite{Landsberg:1985gaz}.
In addition, these Dalitz decays also play an important role in the evaluation of the hadronic light-by-light contribution to the muon anomalous magnetic moment~\cite{Aoyama:2020ynm}.

If one assumes pointlike particles, the electromagnetic decay rate can be precisely calculated by QED. Modifications to the QED decay rate due to the inner structure of the mesons are incorporated in the transition form factor~(TFF) $F(q^2)$, where $q$ is the momentum transferred to the lepton pair~\cite{Landsberg:1985gaz},
\begin{equation}\label{eq1}
    \begin{split}
    &\frac{d\Gamma(P \rightarrow \gamma l^+ l^-)}{dq^2\Gamma(P\rightarrow\gamma\gamma)} \\
    & = \frac{2\alpha}{3\pi}\frac{1}{q^2}\sqrt{1-\frac{4m_l^2}{q^2}}\left (1+\frac{2m_l^2}{q^2} \right)
    \left(1-\frac{q^2}{m^2_{P}} \right)^3|F(q^2)|^2 \\
    & = [{\rm QED}(q^2)]\times|F(q^2)|^2,
    \end{split}
\end{equation}
where $m_{P}$ and $m_l$ are the masses of the pseudoscalar meson ($\eta$ or $\eta'$ in this paper) and the lepton, respectively; $\alpha$ is the fine structure constant; and $[{\rm QED}(q^2)]$ represents the calculable QED part for a pointlike meson. The $F(q^2)$, which is described by phenomenological models, can be experimentally determined from the ratio between the measured dilepton invariant mass spectrum and the QED calculation, which is derived from Eq.~\ref{eq1}. One of the most common phenomenological models to estimate $F(q^2)$ is the vector meson dominance~(VMD) model. In the VMD model~\cite{Sakurai:1969ss}, the interactions between a virtual photon and hadrons are assumed to be mediated through a superposition of neutral vector meson states. Therefore, the TFF is parametrized as~\cite{Landsberg:1985gaz}
\begin{equation}
    F(q^2)=N\sum\limits_{V} \frac{g_{\eta^{(\prime)} \gamma V}}{2g_{V\gamma}}\frac{m^2_V}{m^2_V-q^2-i\Gamma_Vm_V},
\end{equation}
where $N$ is a normalization constant ensuring $F(0)=1$; 
$m_V$ and $\Gamma_V$ are the masses and widths of the vector meson $V=\rho, \omega, \phi$; and $g_{\eta^{(\prime)}\gamma V}$ and $g_{V\gamma}$ are the corresponding coupling constants of the $\eta^{(\prime)}$ transition into a photon and vector meson and the vector meson transition into a photon, respectively.

In a simplified approach, the single-pole form factor is written as
\begin{equation}\label{single-pole}
    F(q^{2})=\frac{1}{1-q^{2}/\Lambda^{2}},
\end{equation}
where the pole parameter $\Lambda$ is expected to be of the order of a vector-meson (typically $\rho, \omega, \phi$) mass. The parameter to be experimentally determined is the slope of the form factor
\begin{equation}
    b=\frac{dF}{dq^2}\bigg|_{q^2=0}=\Lambda^{-2}.
\end{equation}
In the case of the $\eta'$, the pole is expected to lie within the kinematic boundaries of the decay. A widely used expression for the multipole form factor is
\begin{equation}
    \left|F(q^2)\right|^{2}=\frac{\Lambda^2(\Lambda^2+\gamma^2)}{(\Lambda^2-q^2)^2+\Lambda^2\gamma^2},
\end{equation}
where the parameters $\Lambda$ and $\gamma$ correspond to the mass and width of the Breit-Wigner shape for the effective contributing vector meson.
One application of the form factor is to calculate the rms of the interaction regions, $R=\sqrt{6\cdot b}$~\cite{Landsberg:1985gaz}.

In contrast to previous studies of $\eta\to\gamma l^+l^-$ from SND and WASA  \cite{Achasov:2000ne, CELSIUSWASA:2007ifz}, \mbox{BESIII} has unique access to $\eta$ and $\eta^\prime$ decays due to their high production rate in $J/\psi$ radiative and hadronic decays. For example, \mbox{BESIII} reported the first measurement of the $e^+e^-$ invariant-mass distribution from  $\eta^\prime\rightarrow \gamma e^+e^-$ with $J/\psi\rightarrow\gamma\eta'$~\cite{BESIII:2015zpz}. The single-pole parametrization provides a good description of data and the corresponding slope parameter is
$b_{\eta'}=(1.60\pm0.19)$~GeV$^{-2}$. It is in agreement with predictions from different theoretical models~\cite{Bramon:1981sw,Ametller:1983ec,Ametller:1991jv,Hanhart:2013vba} and a previous measurement of 
 $\eta'\to \gamma \mu^+\mu^-$~\cite{Dzhelyadin:1979za}. 

Additionally, new light hidden particles, such as axionlike particles and the dark photon, which may couple with light quarks and gluons, could be produced in the Dalitz decays of pseudoscalar mesons, $\eta/\eta'\rightarrow\gamma e^{+}e^{-}$~\cite{Gan:2020aco}. The dark photon~($A'$) is a new type of force carrier in the simplest scenario of an interaction with dark matter particles and is charged under Abelian $U(1)$ groups~\cite{Arkani-Hamed:2008hhe,Finkbeiner:2007kk, Pospelov:2008zw}. The predicted branching fraction of the meson decay to dark photon is expressed as~\cite{Batell:2009di}
\begin{equation}
    \mathcal{B}(P\rightarrow\gamma A')=2\varepsilon^{2}\left( 1-\frac{m_{A'}^{2}}{m_{P}^{2}} \right)^{3}\mathcal{B}(P\rightarrow\gamma\gamma),
\end{equation}
where $\varepsilon$ is the coupling strength and $m_{A'}$ is the mass of the dark photon. 
An experimental search was performed in $\pi^0\rightarrow\gamma e^{+}e^{-}$ by NA48/2 with a high statistics $\pi^0$ sample tagged in kaon decays~\cite{Raggi:2017gwd}. 
BESIII reported searches for the dark photon in the processes $J/\psi\rightarrow e^{+}e^{-}\eta$~\cite{BESIII:2018qzg} and $J/\psi\rightarrow e^{+}e^{-}\eta'$~\cite{BESIII:2018aao} using a large $J/\psi$ data sample.
However, no search for dark sectors in the Dalitz decays of $\eta$ and $\eta^\prime$ has yet been performed.
 
BESIII has presently collected a sample of 10 billion $J/\psi$ decays~\cite{BESIII:2021cxx}, which is about 8 times larger than that used in Ref.~\cite{BESIII:2015zpz}. This unprecedented sample size allows for improved measurements of the Dalitz decays $\eta/\eta^\prime\rightarrow\gamma e^+e^-$ and searches for new light hidden particles. 

%####################################     event selection
\section{Detector and MC simulation}
The BESIII detector~\cite{BESIII:2009fln} records symmetric $e^+e^-$ collisions provided by the BEPCII storage ring~\cite{Yu:2016cof} in the center-of-mass energy range from 2.0 to 4.95~GeV, with a peak luminosity of $1.1 \times 10^{33}\;\text{cm}^{-2}\text{s}^{-1}$ achieved at $\sqrt{s} = 3.77\;\text{GeV}$. BESIII has collected large data samples in this energy region~\cite{BESIII:2020nme, Jiao:2020dqs, Zhang:2022bdc, Huang:2022wuo}. The cylindrical core of the \mbox{BESIII} detector covers 93\% of the full solid angle and consists of a helium-based multilayer drift chamber~(MDC), a plastic scintillator time-of-flight~(TOF) system, and a CsI(Tl) electromagnetic calorimeter~(EMC), which are all enclosed in a superconducting solenoidal magnet
providing a 1.0~T  (0.9 T in 2012) magnetic field. The solenoid is supported by an octagonal flux-return yoke with resistive plate counter muon identification modules interleaved with steel. 
The charged-particle momentum resolution at $1~{\rm GeV}/c$ is $0.5\%$, and the ${\rm d}E/{\rm d}x$
resolution is $6\%$ for electrons from Bhabha scattering. The EMC measures photon energies with a resolution of $2.5\%$ ($5\%$) at $1$~GeV in the barrel (end cap) region. The time resolution in the TOF barrel region is 68~ps, while that in the end cap region is 110~ps. The end cap TOF system was upgraded in 2015 using multigap resistive plate chamber technology, providing a time resolution of 60 ps~\cite{Cao:2020ibk}.

Simulated data samples produced with a {\sc geant}4-based~\cite{GEANT4:2002zbu} Monte Carlo (MC) package, which includes the geometric description of the BESIII detector and the detector response, are used to determine detection efficiencies and to estimate backgrounds. The simulation models the beam energy spread and initial state radiation in the $e^+e^-$ annihilations with the generator {\sc kkmc}~\cite{Jadach:2000ir}. The inclusive MC sample includes both the production of the $J/\psi$ resonance and the continuum processes incorporated in {\sc kkmc}~\cite{Jadach:2000ir}. 
All particle decays are modeled with {\sc evtgen}~\cite{Lange:2001uf} using branching fractions either taken from the Particle Data Group~(PDG)~\cite{ParticleDataGroup:2022pth}, when available, or otherwise estimated with {\sc lundcharm}~\cite{Chen:2000tv, Yang:2014vra}.
A custom generator, which incorporates theoretical amplitudes, was developed to simulate a variety of exclusive decays, such as $\eta/\eta'\rightarrow\gamma e^{+}e^{-}$~\cite{Qin:2017vkw}, $\eta/\eta'\rightarrow\gamma\pi^{+}\pi^{-}$~\cite{BESIII:2017kyd}, and $J/\psi\rightarrow e^{+}e^{-}\eta/\eta'$~\cite{Gu:2019qwo}.

%%%%%%%%%%%%%%%%%%%%
\section{Event selection} 
Charged tracks are reconstructed from hits in the MDC, and are required to pass within $10$ cm of the interaction point (IP) along the beam direction and within $1$ cm in the plane perpendicular to the beams. 
The polar angle $\theta$ of each charged track is required to be in the range of $\lvert\cos\theta\rvert<0.93$. Photon candidates are reconstructed using clusters in the EMC, where a minimum energy of 0.025 GeV in the barrel region ($\lvert\cos\theta\rvert<0.8$) and 0.050 GeV in the end cap region ($0.86<\lvert\cos\theta\rvert<0.92$) is required. 
Clusters due to electronic noise and deposited energy unrelated to the event are suppressed by requiring the shower time to be within 700 ns of the event start time. To exclude showers produced by charged particles, the angle subtended by the EMC shower and the position of the closest charged track at the EMC must
be greater than $10^\circ$ as measured from the IP.
To study $\eta/\eta'\rightarrow\gamma e^{+}e^{-}$ with $J/\psi\rightarrow\gamma\eta/\eta'$, two oppositely charged tracks and at least two good photons are selected for further analysis.
 
Candidate events are required to successfully pass a primary vertex fit and the two charged tracks are identified as an electron and positron by using combined information from the TOF and ${\rm d}E/{\rm d}x$. Combined likelihoods ($\mathcal{L}$) under the electron and pion hypotheses are obtained. Positron candidates are required to satisfy $\mathcal{L}(e)/(\mathcal{L}(e)+\mathcal{L}(\pi)) > 0.5$ for the $\eta$ and $\mathcal{L}(e)/(\mathcal{L}(e)+\mathcal{L}(\pi)) > 0.95$ for the $\eta'$.
A kinematic 4C fit is then performed under the hypothesis of $J/\psi \to \gamma \gamma e^+ e^-$ by constraining the total four-momentum of the final particles to the initial four-momentum of the $e^+ e^-$ system, and the resulting $\chi^2_{\rm 4C}$ is required to be less than 100. 
For events with more than two photon candidates, the $\gamma\gamma e^+e^-$ combination with minimum $\chi^2_{\rm 4C}$ is retained. Since the radiative photon coming directly from the $J/\psi$ decay is monoenergetic, and has a greater energy than the photon from the $\eta/\eta^\prime$ decays, the photon with maximum energy is regarded as the radiative photon from the $J/\psi$. 

To suppress events from QED processes (i.e., Bhabha events), the energy of the low-energy photon is required to be greater than 0.15 GeV and the angle at the vertex between it and the electron or positron is required to be larger than $10^{\circ}$. To remove background events from $J/\psi$ decays with 
final state radiation photons, in particular  $J/\psi\rightarrow e^+e^-$, the angle between the high-energy radiative photon and the electron or positron is required to be greater than $20^{\circ}$. In addition, the $\gamma\gamma$ invariant mass is required to be outside the mass regions of the $\eta$ and $\eta^\prime$, $|M(\gamma\gamma)-0.547|>0.03~{\rm GeV}/c^{2}$ and $|M(\gamma\gamma)-0.958|>0.03~{\rm GeV}/c^{2}$. This suppresses background events from $J/\psi\rightarrow e^+e^-\eta/\eta^\prime$ with $\eta/\eta'\rightarrow\gamma\gamma$.

\begin{figure}[htbp]
    \centering
    \includegraphics[scale=0.3]{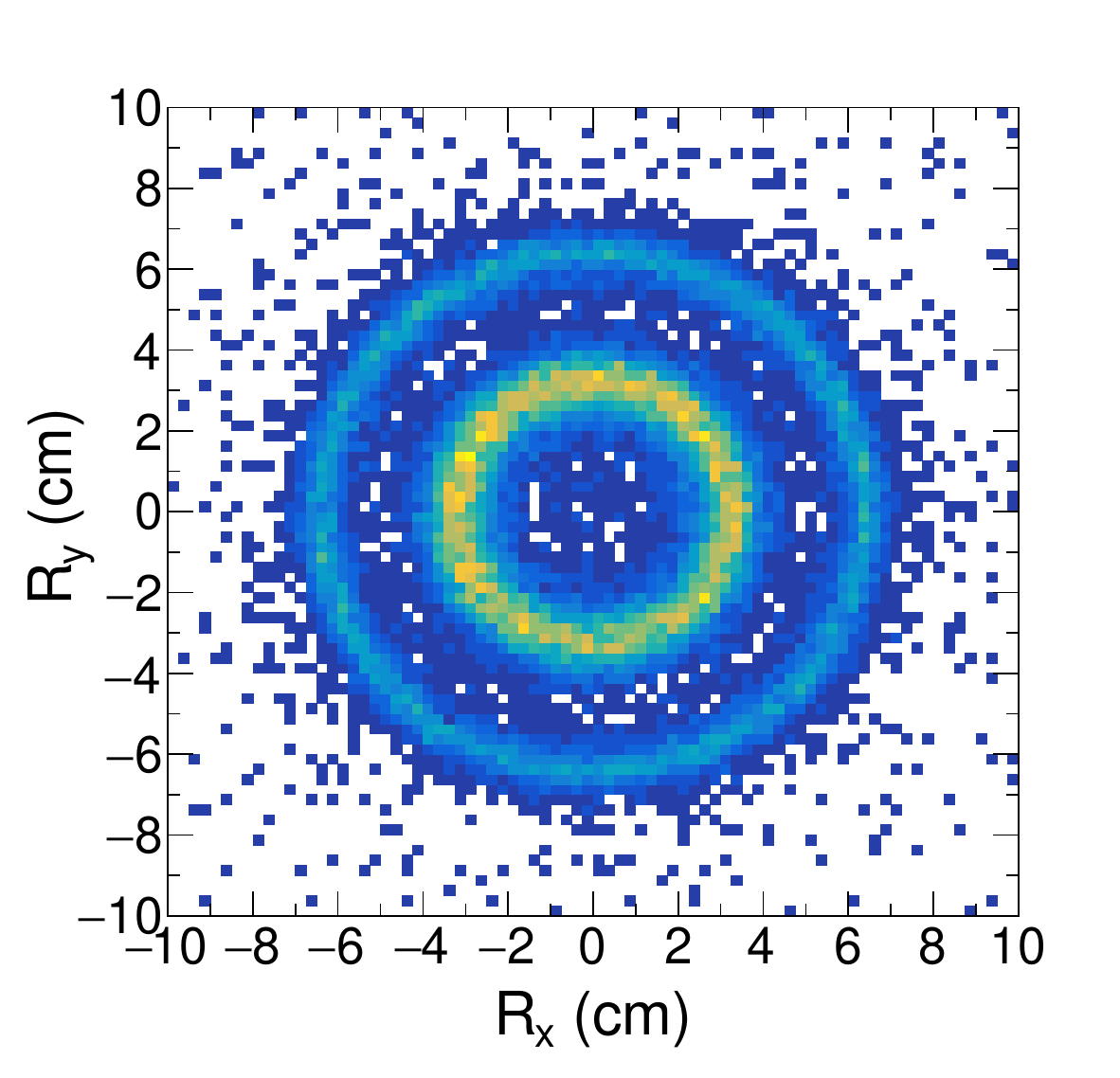}
    \caption{The distribution of the photon conversion points for simulated events from the process $J/\psi\rightarrow\gamma\eta', \eta'\rightarrow\gamma\gamma$ where a photon converts into an electron-positron pair in the beam pipe or inner wall of the MDC.}
    \label{etap:conversion2D}
\end{figure}

\begin{figure}[htbp]
    \centering
    \includegraphics[scale=0.35]{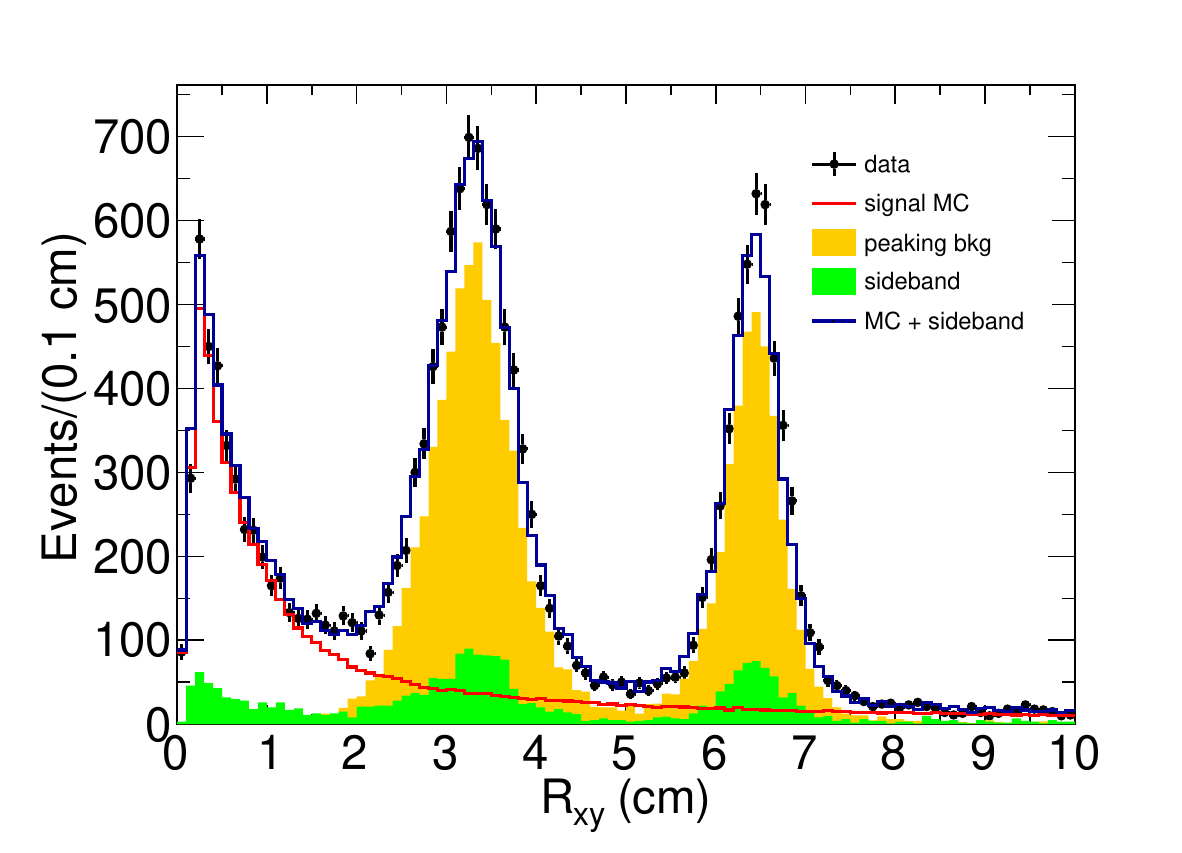}
    \caption{The $R_{xy}$ distributions. The black dots with error bars are data. The red line shows simulated events from the signal process $J/\psi\rightarrow\gamma\eta', \eta'\rightarrow\gamma e^+e^-$. The yellow shaded area shows the background from photon conversion events. The green shaded area is estimated from the $\eta'$ sideband. The blue line is the sum of the simulated signal, the simulated background, and the $\eta'$ sideband background.}
    \label{etap:Rxy}
\end{figure}

After the above requirements, the dominant remaining background contribution is from events where a photon converts into a $e^{+}e^{-}$ pair. To exclude these background events, an algorithm to find
photon conversions is applied to selected $e^+e^-$ pairs. The photon conversion point (CP) is reconstructed using the intersection point of the two charged track trajectories in the $x$-$y$ plane, which is perpendicular to the beam line. The photon conversion length $R_{xy}$ is defined as
the distance from the beam line to the CP in the $x$-$y$ plane along the beam direction. 
Taking $J/\psi\rightarrow \gamma\eta^\prime,~\eta'\rightarrow\gamma\gamma$ as an example, the photon conversion events accumulate at  $R_{xy}\approx3~{\rm cm}$  and $R_{xy}\approx6~{\rm cm}$, which correspond to the positions of the beam pipe and the inner wall of the MDC, respectively, as displayed in Fig.~\ref{etap:conversion2D}. 
The $R_{xy}$ distribution is well reproduced by the MC simulation, as presented in Fig.~\ref{etap:Rxy}.
The signal regions in the invariant mass of $\gamma e^{+}e^{-}$ are defined as $M(\gamma e^{+}e^{-})\in[0.52, 0.56]~({\rm GeV}/c^{2})$ for the $\eta$ and $M(\gamma e^{+}e^{-})\in[0.94, 0.98]~({\rm GeV}/c^{2})$ for the $\eta'$, and the mass sidebands are defined as $M(\gamma e^{+}e^{-})\in[0.50, 0.51]\cup[0.59, 0.60]~{\rm GeV}/c^{2}$ for the $\eta$ and $M(\gamma e^{+}e^{-})\in[0.88, 0.90]\cup[1.00, 1.02]~{\rm GeV}/c^{2}$ for the $\eta'$. 

\begin{figure*}[htb]
    \centering
    \subfigure{
    \begin{overpic}[scale=0.35]{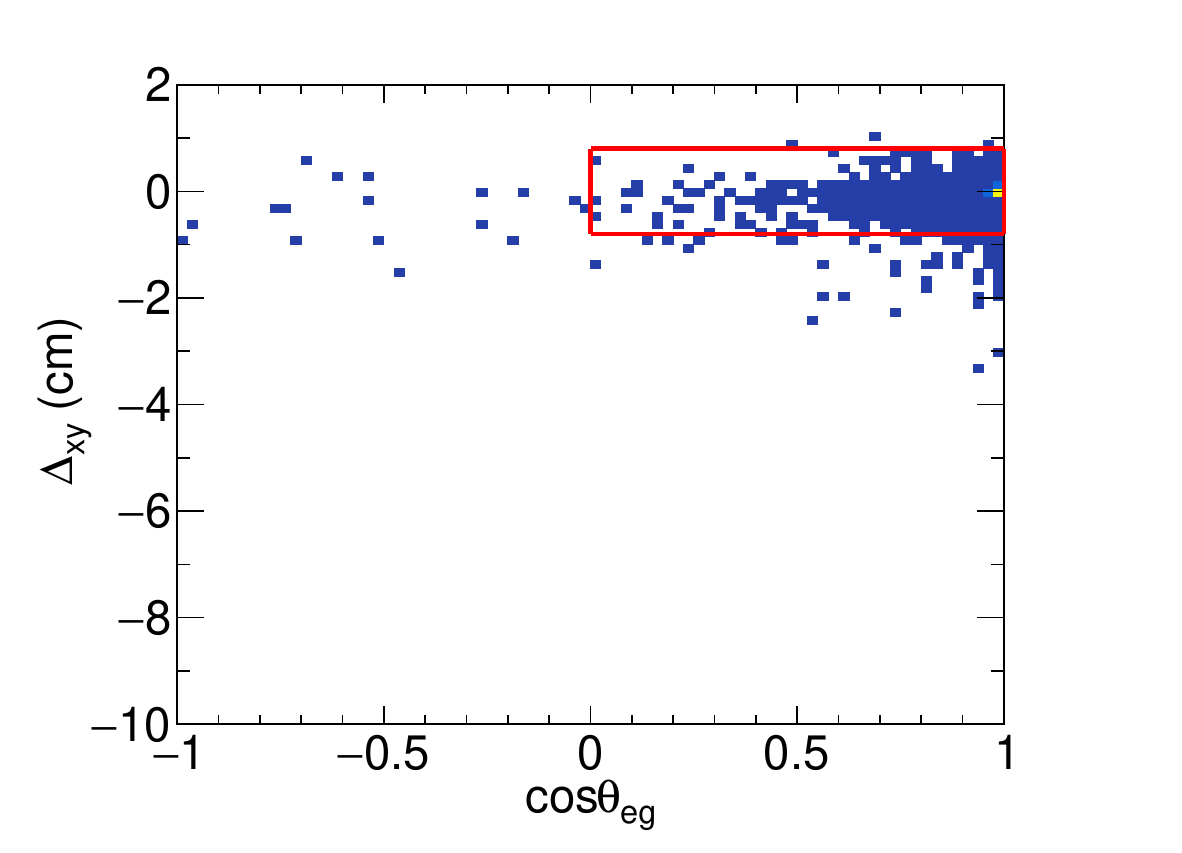}
    \put(46,30){\large (a)}
    \end{overpic}
    }
    \subfigure{
    \begin{overpic}[scale=0.35]{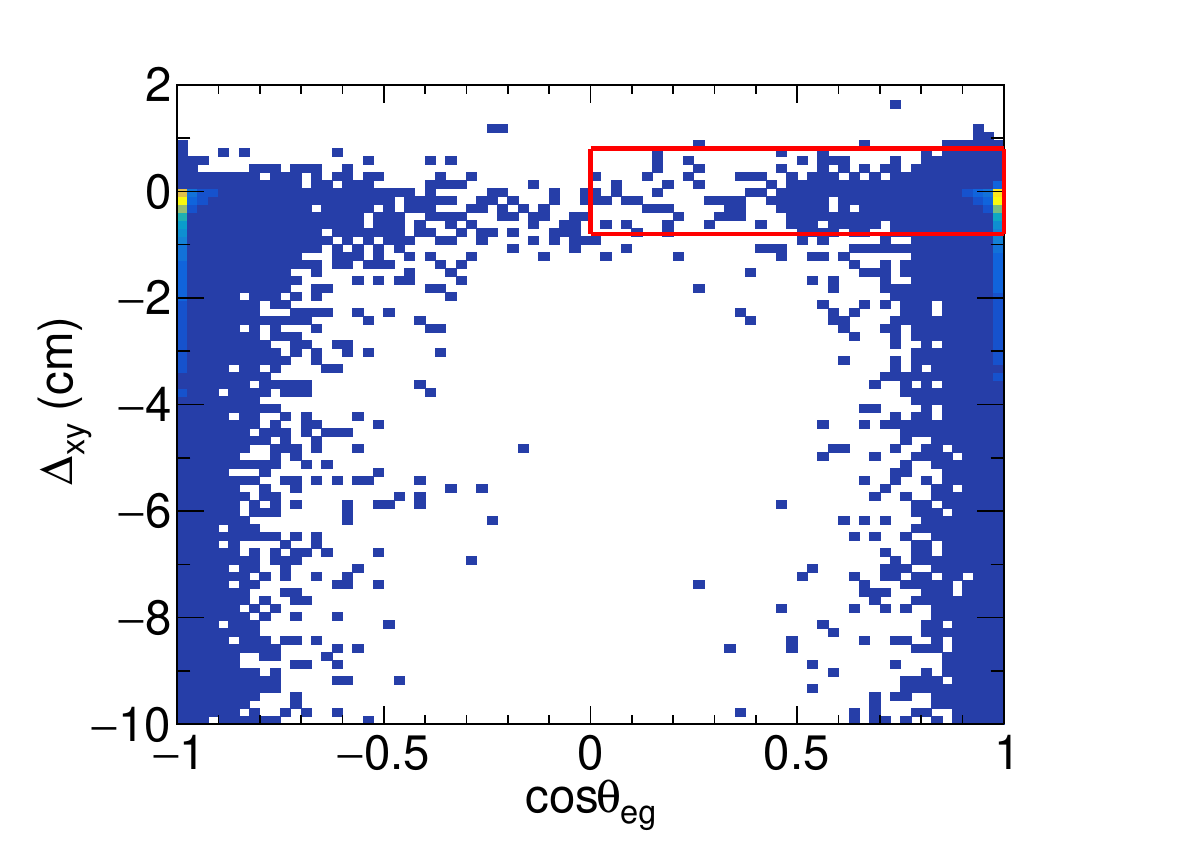}
    \put(46,30){\large (b)}
    \end{overpic}
    }
    \caption{The distributions of $\Delta_{xy}$ versus $\cos{\theta_{eg}}$ for simulated events with $R_{xy}>2~{\rm cm}$ from the processes (a) $J/\psi\rightarrow\gamma\eta', \eta'\rightarrow\gamma\gamma$ and (b) $J/\psi\rightarrow\gamma\eta', \eta'\rightarrow\gamma e^{+}e^{-}$. The rectangle, which corresponds to $|\Delta_{xy}|<0.8~{\rm cm}$ and $\cos\theta_{eg}>0$, is the rejection range for photon conversion events.}
    \label{etap:cosvsdelta}
\end{figure*}

To preserve more signal events, we only reject events with $R_{xy}>2~{\rm cm}$ when $\cos{\theta_{eg}}>0$ and $|\Delta_{xy}|<0.8~{\rm cm}$. Here, $\cos{\theta_{eg}}$ is the angle between the momentum vector of the reconstructed photon, which converts to an electron-positron pair, and the direction from the IP to the conversion point. The electron and positron tracks project to two circles in the $x$-$y$ plane, and we define $\Delta_{xy}$ as the distance between the intersections of the two circles and the line connecting the centers of the circles.
The distributions of $\Delta_{xy}$ versus $\cos\theta_{eg}$ from the simulated events are displayed in Fig.~\ref{etap:cosvsdelta}.

After imposing the additional requirements discussed above, Fig.~\ref{eta:etafit} shows the invariant mass ($M(\gamma e^+e^-)$) spectra in the $\eta$ and $\eta^\prime$ mass regions, respectively, where clear $\eta$ and $\eta^\prime$ peaks are observed. To estimate the remaining background contributions, we have performed extensive studies of potential background processes using data taken at the center-of-mass energy of 3.08 GeV and the simulated inclusive $\jpsi$ sample.
The results indicate that the nonpeaking background events mainly come from the processes $e^+e^-\rightarrow e^+e^- \gamma$, $e^+e^-\rightarrow e^+e^- \gamma\gamma$, $e^+e^-\rightarrow \gamma\gamma\gamma$, $J/\psi\rightarrow e^{+}e^{-}$, and $J/\psi\rightarrow\gamma\pi^0\pi^0$.
Using dedicated MC samples for $J/\psi\rightarrow \gamma \eta/\eta^\prime$ with $\eta/\eta'\rightarrow\gamma\gamma$, the numbers of normalized peaking background events are determined to be $729 \pm 26$ and $192 \pm 12$, respectively, in accordance with the corresponding branching fractions in the PDG~\cite{ParticleDataGroup:2022pth}. The quoted errors are due to the MC statistics and the uncertainties of branching fractions.

\section{Signal yields of $ \eta/\eta^\prime\rightarrow \gamma e^+e^-$}

The signal yields for the decays $\eta/\eta^\prime \rightarrow \gamma e^+e^-$ are obtained from extended unbinned maximum likelihood fits to the $M(\gamma e^+e^-)$ distributions. The total probability density function consists of a signal and various background contributions.
The signal component is modeled by the MC-simulated signal shape convolved with a  Gaussian function to account for the difference in the mass resolution
between data and MC simulation. The background components are subdivided into two classes: (i) the shapes of photon conversion events are taken from the dedicated MC simulations and the magnitudes are fixed as described above; (ii) the continuum background events are described by a first-order Chebychev function. 

\begin{figure*}[htbp]
    \centering
    \subfigure{
    \begin{overpic}[scale=0.35]{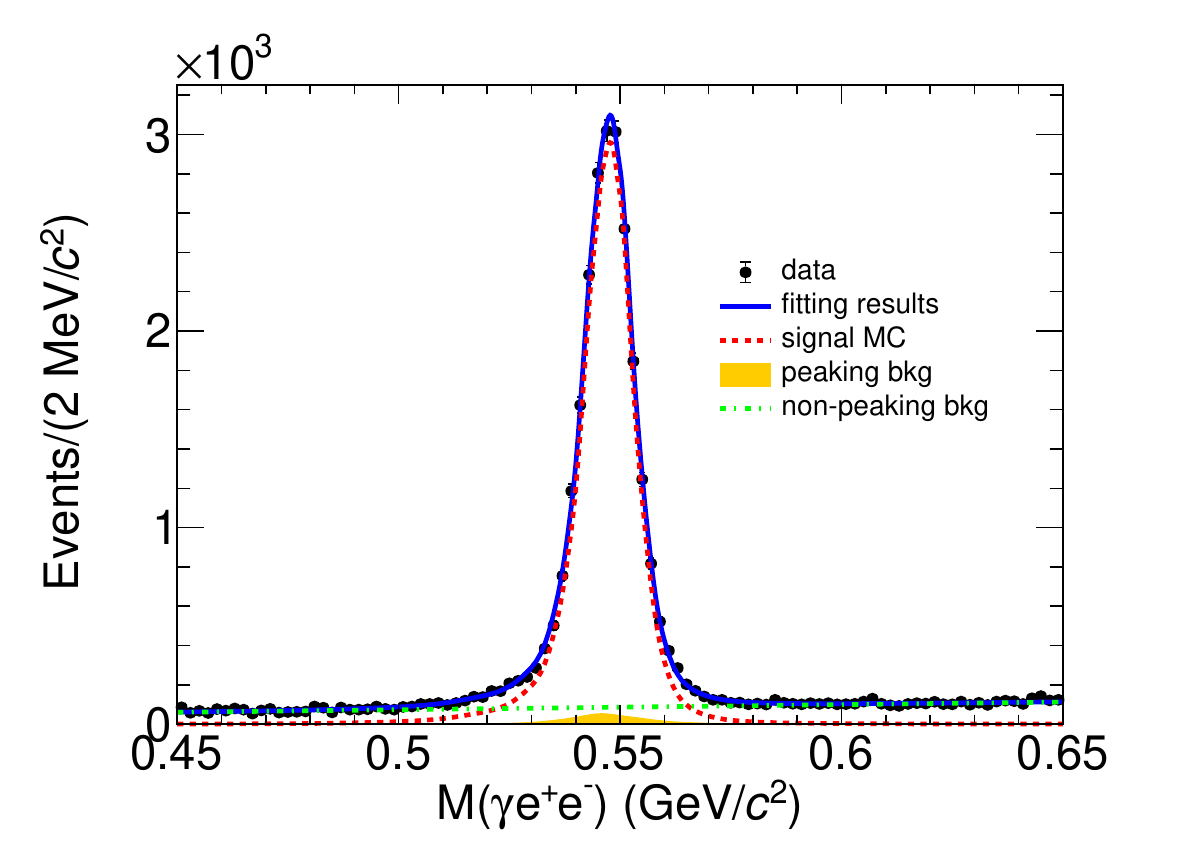}
    \put(20,55){\large (a)}
    \end{overpic}
    }
    \subfigure{
    \begin{overpic}[scale=0.35]{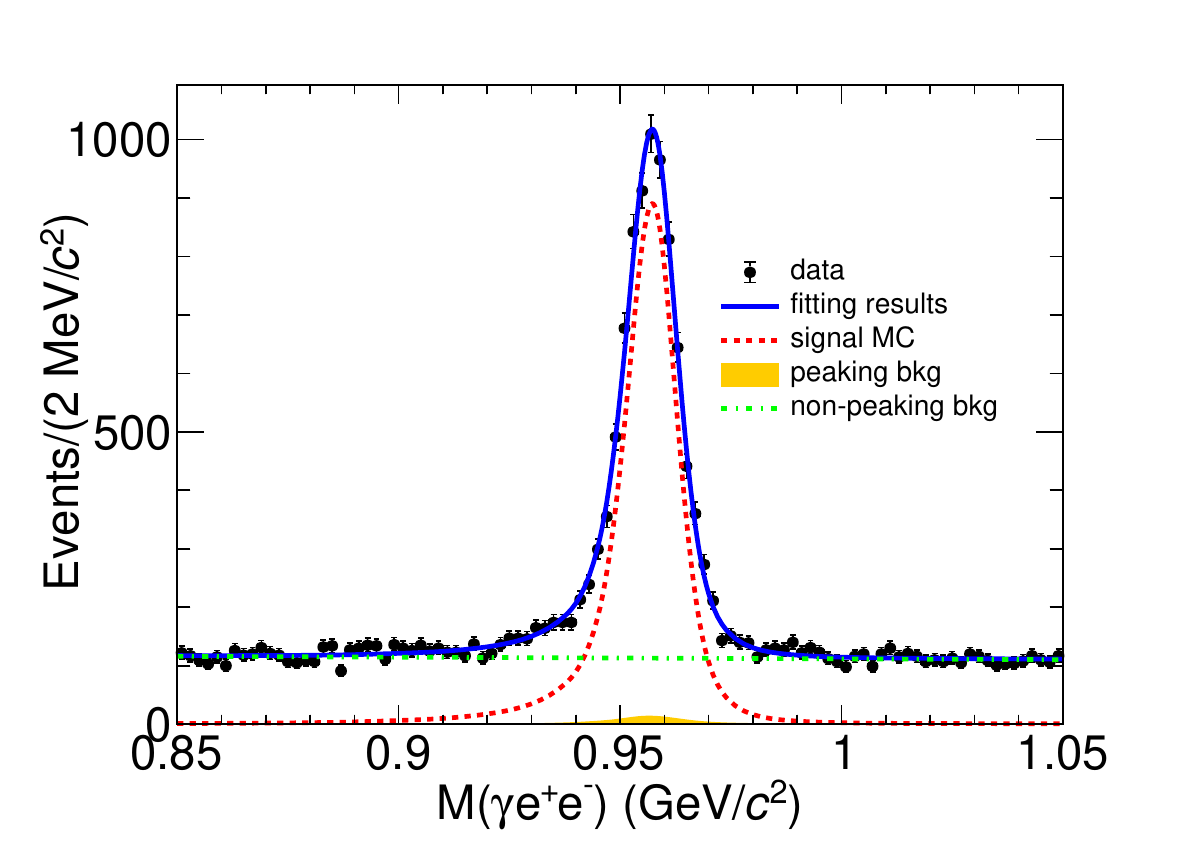}
    \put(20,55){\large (b)}
    \end{overpic}
    }
    \caption{The fits to the $M(\gamma e^{+}e^{-})$ distributions for (a) $\eta\rightarrow\gamma e^{+}e^{-}$ and (b) $\eta'\rightarrow\gamma e^{+}e^{-}$. The dots with error bars represent data, the blue line represents the total fit, the red dotted line represents the signal, the yellow shaded area represents the peaking background from photon conversion, and the green line represents the fitted Chebychev function.}
    \label{eta:etafit}
\end{figure*}

The fits yield $22907\pm164$  $\eta\rightarrow\gamma e^+e^-$ events and 
$7611\pm 108$ $\eta^\prime\rightarrow\gamma e^+e^-$ events. 
The results of the fits to the $M(\gamma e^+e^-)$ distributions in the $\eta$ and $\eta^\prime$ mass regions are shown in Fig.~\ref{eta:etafit}.

Using the measured signal yields, the branching fractions are determined with the following equations:
\begin{equation}
    \mathcal{B}(\eta/\eta'\rightarrow\gamma e^+e^-)=\frac{ N_{\rm obs} }{ N_{J/\psi}\cdot\epsilon\cdot\mathcal{B}(J/\psi\rightarrow\gamma\eta/\eta') }
\end{equation}
where $N_{\rm obs}$ is the observed signal events, $N_{J/\psi}$ is the number of $J/\psi$ events, $\epsilon$ is the detection efficiency and $\mathcal{B}(J/\psi\rightarrow\gamma\eta/\eta')$ is the branching fraction referred from PDG~\cite{ParticleDataGroup:2022pth}. With a detection efficiency of $(29.60\pm0.03)\%$ for $\eta$ and $(29.80\pm0.03)\%$ for $\eta'$, the branching fraction of $\eta/\eta'\to\gamma e^{+}e^{-}$ is calculated to be $(7.07\pm0.05)\times 10^{-3}$ and $(4.83\pm0.07)\times10^{-4}$ with statistical uncertainties, respectively.

\section{Form factor measurement}
\label{section:this}
To extract the electromagnetic transition form factor of the $\eta/\eta'$, an unbinned maximum likelihood fit to the $M(e^+e^-)$ spectrum in data is performed. To measure the $\eta / \eta'$ form factor, $M(\gamma e^+e^-)$ is required to be in the $\eta/\eta'$ signal regions which are defined as $M(\gamma e^{+}e^{-})\in[0.52, 0.56]~({\rm GeV}/c^{2})$ for the $\eta$ and $M(\gamma e^{+}e^{-})\in[0.94, 0.98]~({\rm GeV}/c^{2})$ for the $\eta'$, and the mass sidebands, which are defined as $M(\gamma e^{+}e^{-})\in[0.50, 0.51]\cup[0.59, 0.60]~{\rm GeV}/c^{2}$ for the $\eta$ and $M(\gamma e^{+}e^{-})\in[0.88, 0.90]\cup[1.00, 1.02]~{\rm GeV}/c^{2}$ for the $\eta'$, are used to estimate the nonpeaking background contributions. 
The peaking background contribution from photon conversion events is estimated from the MC simulation, and the branching fractions are taken from the PDG~\cite{ParticleDataGroup:2022pth}. The corresponding  background yields from $J/\psi\rightarrow\gamma\eta,~\eta\rightarrow\gamma\gamma$ and $J/\psi\rightarrow\gamma\eta',~\eta'\rightarrow\gamma\gamma$ are estimated to be $582\pm21$ and $155\pm9$, respectively.
The numbers of candidates selected in the $\eta$ and $\eta^\prime$ mass regions are 23189 and 9333, respectively.

The free parameter of the probability-density function (PDF) to observe the $i$ th event characterized by the measured four-momenta $\xi_{i}$ of the particles in the final state is
\begin{equation}
    \mathcal{P}(\xi_{i})=\frac{|\mathcal{A}(\xi_{i})|^{2} \epsilon(\xi_{i})}{\int d\xi |A(\xi)|^{2}\epsilon(\xi)},
\end{equation}
where $\mathcal{A}$ is the full amplitude including all the potential intermediate states in Eq.~(\ref{eq1}), and $\epsilon(\xi_{i})$ is the detection efficiency. The free parameters are estimated by \mbox{MINUIT}~\cite{James:1975dr}. The fit minimizes the negative log-likelihood value
\begin{equation}
\begin{split}
    &-\ln{\mathcal{L}}=-\omega^{\prime}\Bigg[ \sum^{N_{data}}_{i=1}\ln{\mathcal{P}(\xi_{i})} \\ &-\omega_{bkg1}\sum^{N_{bkg1}}_{j=1}\ln{\mathcal{P}(\xi_{j})} -\omega_{bkg2}\sum^{N_{bkg2}}_{k=1}\ln{\mathcal{P}(\xi_{k})} \Bigg],
\end{split}
\end{equation}
where $\mathcal{P}$ is the PDF, $i$, $j$ and $k$ run over all accepted data, the peaking background, and the continuous background events, respectively, and their corresponding number of events are denoted by $N_{data}$, $N_{bkg1}$, and $N_{bkg2}$. Here, $\omega_{bkg1}=\frac{N^{\prime}_{bkg1}}{N_{bkg1}}$ and $\omega_{bkg2}=\frac{N^{\prime}_{bkg2}}{N_{bkg2}}$ are the weights of the backgrounds, where $N^{\prime}_{bkg1}$ and $N^{\prime}_{bkg2}$ are their contributions in the signal region according to branching fractions taken from PDG~\cite{ParticleDataGroup:2022pth} and the above fitting results as displayed in Fig.~\ref{eta:etafit}. To obtain an unbiased uncertainty estimation, the normalization factor derived from Ref.~\cite{Langenbruch:2019nwe} is considered, described as
\begin{equation}
    \omega^{\prime}=\frac{ N_{data}-N_{bkg1}\omega_{bkg1}-N_{bkg2}\omega_{bkg2} }{ N_{data}+N_{bkg1}\omega_{bkg1}^{2}+N_{bkg2}\omega_{bkg2}^{2} }.
\end{equation}

\begin{figure*}[htbp]
    \centering
    \subfigure{
    \begin{overpic}[scale=0.35]{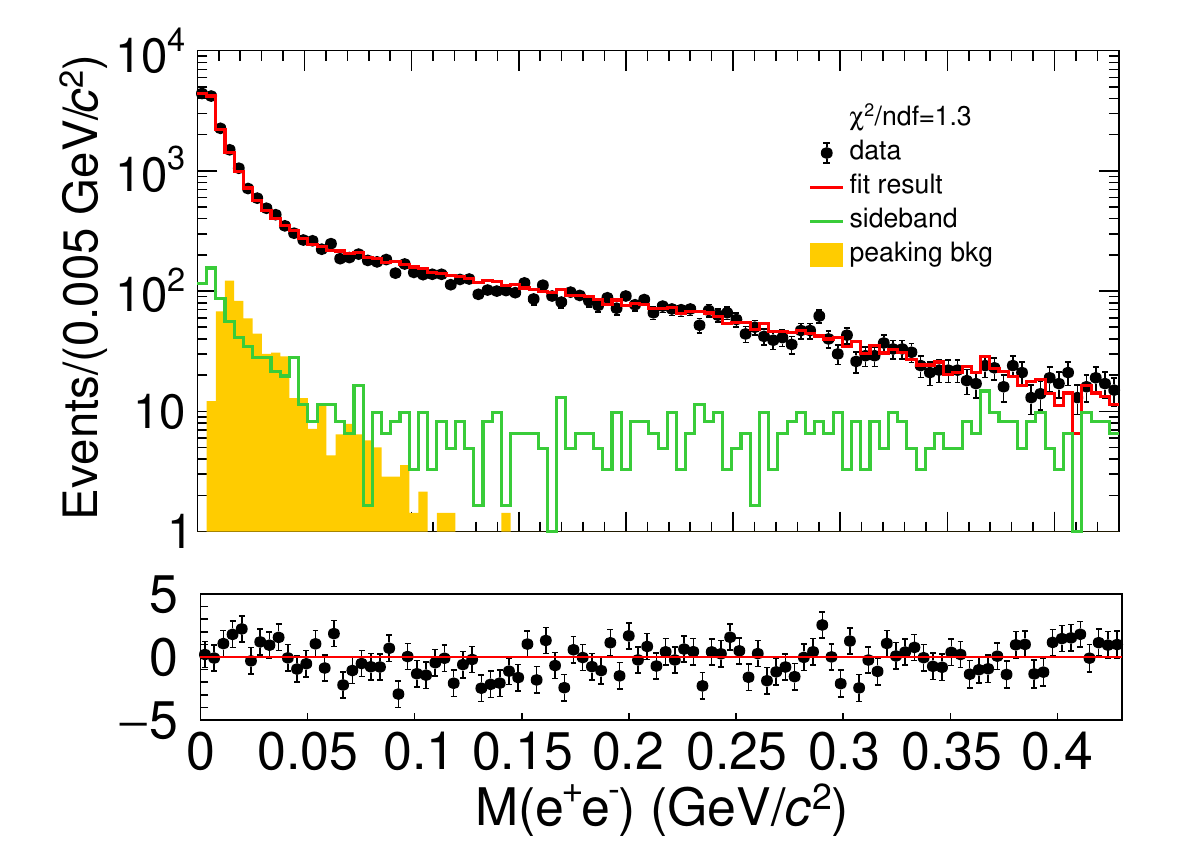}
    \put(20,60){\large (a)}
    \end{overpic}
    \label{eta:tff}
    }
    \subfigure{
    \begin{overpic}[scale=0.35]{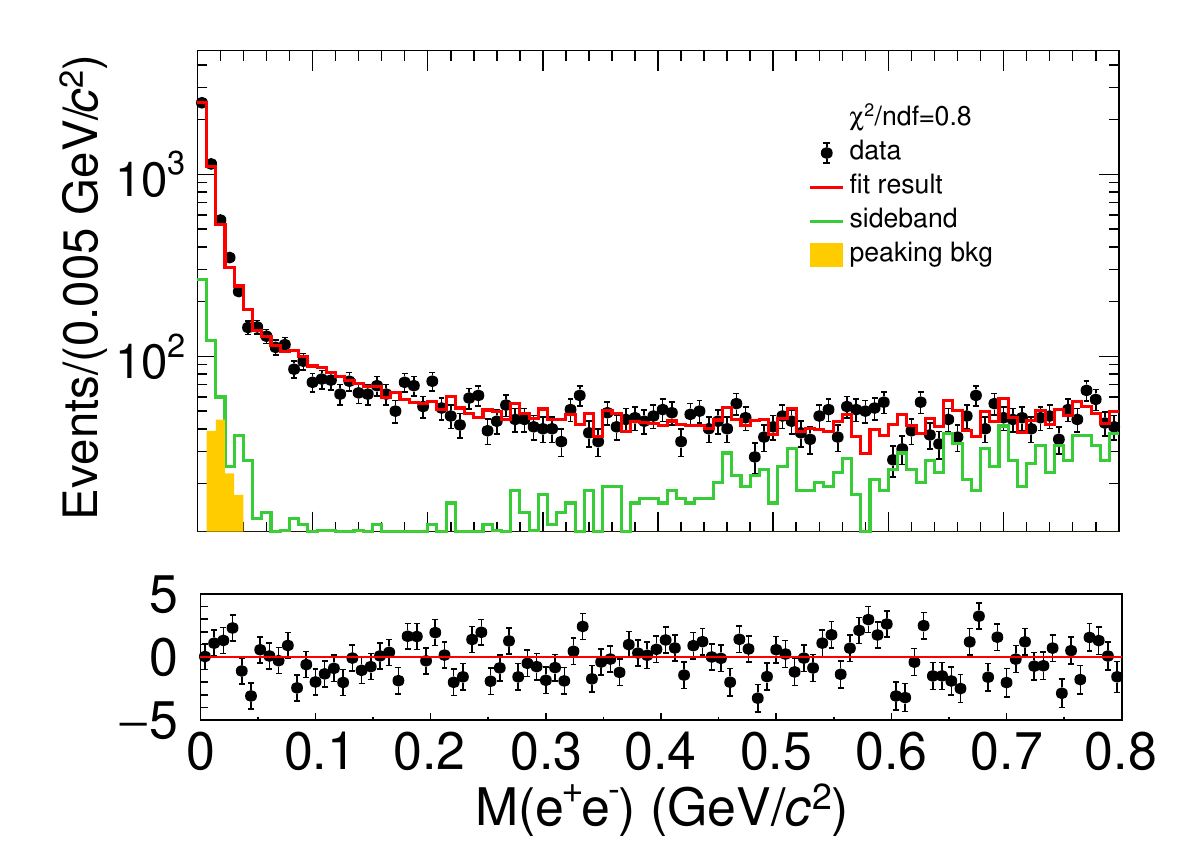}
    \put(20,60){\large (b)}
    \end{overpic}
    \label{etap:tff}
    }
    \caption{The fits to the $\eta^{(\prime)}$ form factors for (a) $\eta\rightarrow\gamma e^{+}e^{-}$ and (b) $\etap\rightarrow\gamma e^{+}e^{-}$. The black dots with error bars represent data, the red histogram represents the total fit, the yellow shaded histogram represents the peaking background from the simulated events of $J/\psi\rightarrow\gamma\eta, \eta\rightarrow\gamma\gamma$, and the green histogram represents the $\eta^{(\prime)}$ sideband.}
\end{figure*}

For the $\eta$ case, we fit the $M(e^+e^-)$ distribution with the single-pole model, and the result is illustrated in Fig.~\ref{eta:tff}. The fit gives $\Lambda_{\eta}=(0.749 \pm 0.027)~{\rm GeV}/c^{2}$. The corresponding radius of the interaction region is calculated to be $R_{\eta}=(0.645 \pm 0.023)~{\rm fm}$~\cite{Landsberg:1985gaz}. Here the uncertainties are statistical only.

For the $\eta^\prime$ case, the single-pole model does not describe the data well. Instead, we fit the $M(e^{+}e^{-})$ distribution with the multipole model, and the result is shown in Fig.~\ref{etap:tff}. From the fit, we obtain $\Lambda_{\eta'}=(0.802 \pm 0.007)~{\rm GeV}/c^{2}$ and $\gamma_{\eta'}=(0.113\pm0.010)~{\rm GeV}/c^{2}$. The radius of the interaction region is calculated to be $R_{\eta'}=(0.596 \pm 0.005)~{\rm fm}$~\cite{Landsberg:1985gaz}.  The uncertainties are statistical only.

\section{Systematic uncertainty}

In the measurements of the branching fractions and the electromagnetic form factors, the possible sources of systematic uncertainties are summarized in Tables~\ref{tab:totalerr} and~\ref{tab:tfferr}, as discussed in detail below.

\subsection{MDC tracking}
The MDC tracking efficiency of electrons has been studied with a control sample of both radiative Bhabha events, $e^+e^-\rightarrow \gamma e^+e^-$, at the $J/\psi$ peak and $J/\psi\rightarrow e^+e^-$ events. The data-MC difference, $\Delta_{\rm syst.}$, is extracted as a function of the particle momentum and the polar angle. Subsequently, each event in the MC sample is reweighted by a factor (1 + $\Delta_{\rm syst.}$). The branching fractions are recalculated with efficiencies determined from the reweighted MC sample. For the TFF measurement, a reweighted MC sample is used to calculate the MC integral, and a group of new fitting results are obtained using the same fit method as in Sec.~\ref{section:this}. The difference with the nominal result is taken as the systematic uncertainty.

\subsection{PID efficiency}
The PID efficiency of electrons has also been studied with a control sample of both radiative Bhabha events, $e^+e^-\rightarrow \gamma e^+e^-$, at the $J/\psi$ peak and $J/\psi\rightarrow e^+e^-$ events. Using the same approach as that used for the tracking efficiency, we correct the particle identification (PID) efficiency as a function of both $\cos{\theta}$ ranges and the transverse momentum for each electron. The change of the efficiency is assigned as the systematic uncertainty. 

\subsection{Photon detection}
For photons directly detected by the EMC, the detection efficiency has been studied using a control sample of $e^+e^-\rightarrow \gamma_{ISR} \mu^+\mu^- $ events. The four-momentum of the initial-state-radiation photon is predicted using only the four-momentum of the $\mu^+\mu^- $ pair. 
The photon detection efficiency is determined from the fraction of events 
containing a photon with a four-momentum consistent with the prediction. The systematic uncertainty is defined as the relative difference in the efficiency between data and MC simulation. 

\subsection{4C kinematic fit}
The systematic uncertainty associated with the 4C kinematic fit comes from track reconstruction inconsistency between the data and MC simulation. This difference has been reduced by
correcting the track helix parameters of the MC simulation~\cite{BESIII:2012mpj}. 
We take the efficiency with the correction as the nominal value, and take the difference between the efficiencies with and without the correction as the systematic uncertainty.

\subsection{Photon conversion veto}

To evaluate the systematic uncertainty associated with the rejection of photon conversions, we select a clean sample of  $J/\psi\rightarrow\pi^+\pi^-\pi^0, \pi^0\rightarrow \gamma e^+e^-$ events, which includes both  $\pi^0$ Dalitz decays and $\pi^0\to\gamma\gamma$ decays where one photon externally converts to an electron-positron pair.  Using the same requirements to veto photon conversion events as described above, the data-MC difference of the selection efficiencies is evaluated and taken as the systematic uncertainty in the calculation of the branching fractions. 

Using the same approach as was used for the tracking efficiency, we do a two-dimensional correction to the photon conversion veto efficiency as a function of the momentum of the electron and positron. The change of the efficiency is assigned as the systematic uncertainty.

\subsection{$M(\gamma e^+e^-)$ fit}
The uncertainties due to the fit range are considered by varying the fit ranges, and the uncertainty associated with the smooth background function is evaluated by replacing the shape in the nominal fit with a second-order polynomial.

\subsection{Number of $J/\psi$ events}
Using $J/\psi$ inclusive decays, the number of $J/\psi$ events is determined to be $(10087\pm44)\times10^{6}$~\cite{BESIII:2021cxx} and the corresponding uncertainty, 0.44\%, is assigned as the systematic uncertainty in the calculation of the branching fraction.

\subsection{Quoted branching fractions}
External uncertainties from the branching fractions of $J/\psi\rightarrow\gamma\eta/\eta^\prime$ are directly taken from the world average values~\cite{ParticleDataGroup:2022pth}. To account for the form factor uncertainties due to the peaking background contributions from $J/\psi\rightarrow\gamma\eta/\eta^\prime$, $\eta/\eta^\prime\rightarrow\gamma\gamma$, the normalized background events are varied by 1 standard deviation in the alternative fits in accordance with the uncertainties of the branching fractions in the PDG~\cite{ParticleDataGroup:2022pth}. The changes with respect to the nominal results are assigned as systematic uncertainties.

\subsection{Signal model}
In the fit to determine the $\eta/\eta'$ yield, the signal shape is determined from signal MC events which are generated using a custom generator~\cite{Qin:2017vkw}. 
The parameters of the form factors used in the generator are determined with an iterative procedure. 
A new MC sample is generated with parameters determined from the data using the MC sample generated
at the previous step. Already after the first iteration, the changes of selection efficiencies
are less than 0.03\%, and this source of systematic uncertainty is ignored.
We also use a double-sided crystal ball function instead of the signal MC shape to fit the $M(e^{+}e^{-})$ distributions. The systematic uncertainty is determined to be 1.0\% for both the $\eta$ and $\eta'$.

All systematic uncertainties in the branching fraction and form factor measurements are summarized in Tables ~\ref{tab:totalerr} and ~\ref{tab:tfferr}, respectively. 
Assuming all the sources are independent, the total systematic uncertainties are obtained by adding the individual values in quadrature. The obtained results are shown in the last row of each table. 

\begin{table}[htbp]
    \caption{The systematic uncertainties of the measured branching fractions (in percentage).}
    \centering
    \setlength{\tabcolsep}{3mm}{
    %\resizebox{\textwidth}{100mm}{
    \renewcommand{\arraystretch}{1.1}
    \begin{tabular}{l|c|c}
    \hline
        Sources & $\eta\rightarrow\gamma e^+e^-$ & $\eta^\prime\rightarrow\gamma e^+e^-$ \\
    \hline
        MDC tracking & 2.2&1.1 \\
        PID & 0.9 & 0.5 \\
        Photon detection &0.5 &  0.6 \\
        Kinematic fit &0.3 &  0.3  \\
       Photon conversion  veto&0.9& 0.9  \\
        $M(\gamma e^{+}e^{-})$ fit  & 0.6& 1.5  \\
        Number of $J/\psi$ events&0.44 & 0.44 \\
        $ \mathcal{B}(J/\psi \rightarrow \gamma \eta/\eta')$ &1.3& 1.3 \\
        Signal shape & 1.0 & 1.0 \\
    \hline
        Total & 3.2 & 2.8 \\
    \hline
    \end{tabular}}
    \label{tab:totalerr}
\end{table}

\begin{table}[htbp]
    \caption{The systematic uncertainties of the measured form factors~(in percentage).}
    \centering
    \setlength{\tabcolsep}{4.3mm}{
    %\resizebox{\textwidth}{3mm}{
    \renewcommand{\arraystretch}{1.1}
    \begin{tabular}{l |c | c c}
    \hline
        Sources & $\Lambda_{\eta}$ & $\Lambda_{\eta'}$ & $\gamma_{\eta'}$\\
    \hline
        MDC tracking  & 0.5 & 0.1 & 0.3 \\
        PID           & 0.5 & 0.1 & 0.1 \\
        Kinematic fit & 0.2 & 0.1 & 0.3 \\
        Background uncertainties & 0.1 & 0.2 & 0.2 \\
        Veto of photon conversion & 0.6 & 1.0 & 1.3 \\
    \hline
        Total & 1.0 & 1.1 & 1.4 \\
    \hline
    \end{tabular}}
    \label{tab:tfferr}
\end{table}

\section{search for the dark photon}

\begin{figure}[htbp]
    \centering
    \includegraphics[scale=0.3]{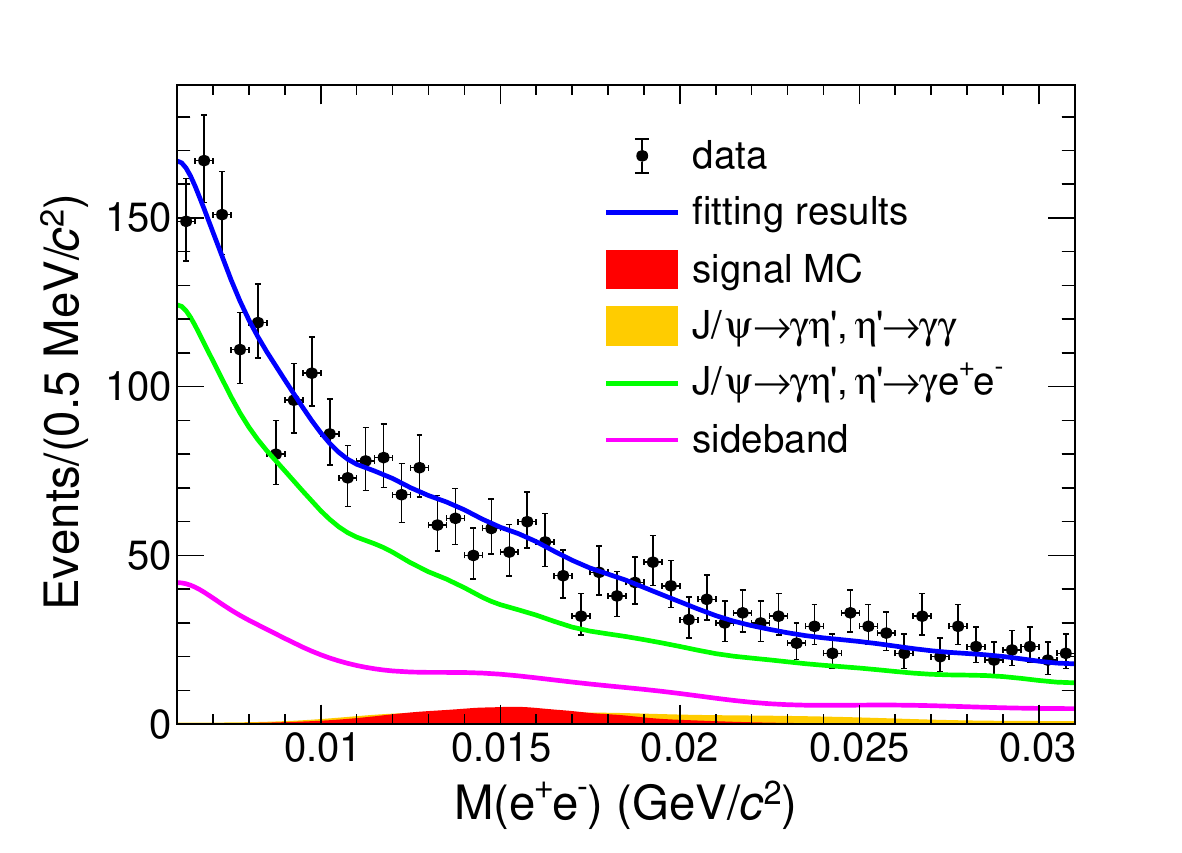}
    \caption{Fit to the invariant-mass distribution of $M(e^{+}e^{-})$ when the invariant mass of $A'$ is $15~{\rm MeV}/c^{2}$. The dots with error bars represent data, and the blue solid line is the total fitting results. The red histogram represents the arbitrary normalized MC signal shape. The green dotted histogram is the $J/\psi\rightarrow\gamma \eta',~\eta'\rightarrow\gamma e^{+}e^{-}$ MC shape. The yellow dotted histogram is the $J/\psi\rightarrow\gamma\eta',~\eta'\rightarrow\gamma\gamma$ MC shape. The gray dotted histogram is the background obtained from the $\eta'$ sideband.}
    \label{Ndp}
\end{figure}

We search for the dark photon $A'$ with $A'\to e^{+}e^{-}$ by 
performing a series of unbinned extended maximum likelihood fits to the $M(e^+e^-)$ distribution.  The mass of the $A^{\prime}$ is scanned over a range of $[0, 0.4]~{\rm GeV}/c^{2}$ for the $\eta$ and $[0, 0.7]~{\rm GeV}/c^{2}$ for the $\eta'$ with a step length of $0.01~{\rm GeV}/c^{2}$.  
The $A^\prime$ signal is described by the MC-simulated shape with an assumption of a negligible width, where $\eta/\eta'\to\gamma A'$ and $A'\to e^{+}e^{-}$ are all modeled by phase space. The background components are subdivided into three classes: (i) the shapes of the photon conversion background events of $\eta/\eta^\prime\rightarrow\gamma\gamma$ that contribute to a structure in the $\eta/\eta^\prime$ mass regions are taken from the dedicated MC simulation; (ii)  the background contribution of the Dalitz decays $\eta/\eta^\prime\rightarrow\gamma e^{+}e^{-}$ is modeled with the above results; (iii) the remaining background events are estimated with the corresponding sidebands,  defined as
$M(\gamma e^{+}e^{-})\in[0.50, 0.51]\cup[0.59, 0.60]~{\rm GeV}/c^{2}$  for $\eta$ and $M(\gamma e^{+}e^{-})\in[0.88, 0.90]\cup[1.00, 1.02]~{\rm GeV}/c^{2}$ for $\eta^\prime$. With MC simulation, no peaking background is considered. 
As an example shown in Fig.~\ref{Ndp}, the significance for each assumed $A^{\prime}$ mass is less than $0.5\sigma$.

Several multiplicative systematic uncertainties for each mass hypothesis of the $\mathbf{\it A'}$ are studied, including: photon detection~(0.5\%), veto of photon conversion~(0.9\%), $N_{J/\psi}$~(0.44\%),  $\mathcal{B}(J/\psi\rightarrow\gamma P)$~(1.3\%), MC statistics~(0.4\%), MDC tracking~(0.7\%-3.1\%), PID~(0.8\%-1.2\%) and kinematic fit~(0.6\%-7.5\%). The total systematic uncertainty is 2\%-8\%. The additive systematic uncertainties are considered by alternative fit ranges and background models. The maximum number of signal events among the different fit scenarios is adopted to calculate the upper limit of the signal yield.

\begin{figure*}[htbp]
    \centering
    \subfigure{
    \begin{overpic}[scale=0.35]{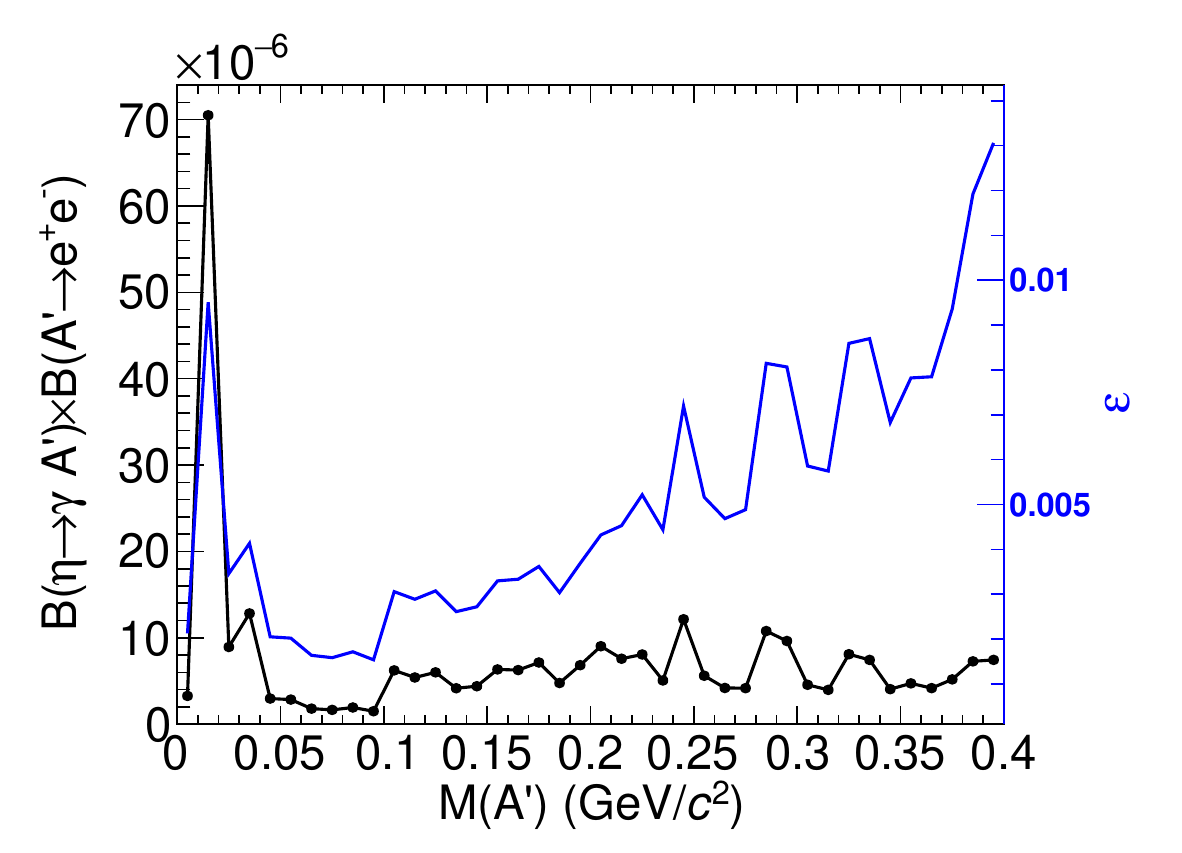}
    \put(20,55){\large (a)}
    \end{overpic}
    }
    \subfigure{
    \begin{overpic}[scale=0.35]{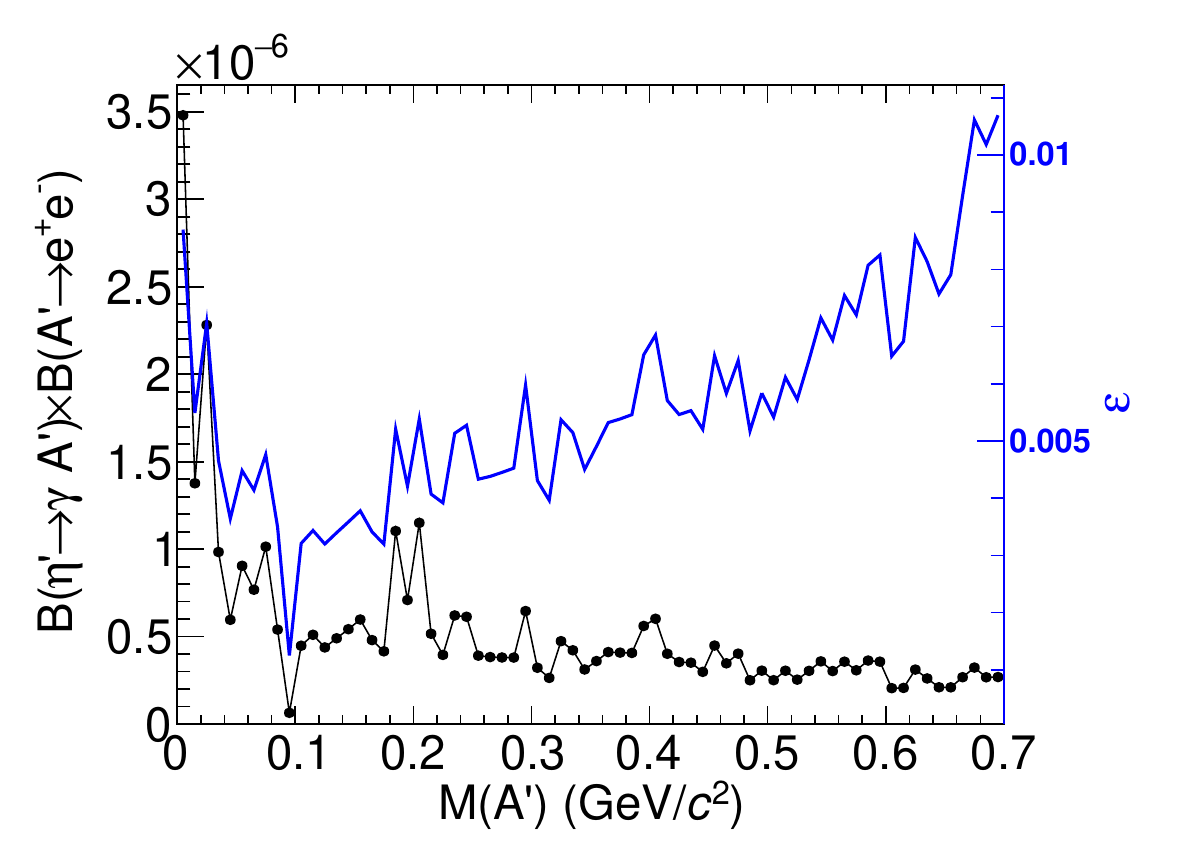}
    \put(20,55){\large (b)}
    \end{overpic}
    }
    \caption{The upper limit of the branching fractions of the dark photon (shown as the black line) and the coupling strength $\varepsilon$ (shown as the blue line) for (a) $\eta\rightarrow\gamma A'$ and (b) $\eta'\rightarrow\gamma A'$.}
    \label{aeta}
\end{figure*}

Since no $A^\prime$ signal is apparent in the $M(e^+e^-)$ distribution, we compute
the upper limit on the product branching fraction $\mathcal{B}(P\rightarrow \gamma A^\prime)\times\mathcal{B}(A^\prime\rightarrow e^+e^-)$ at the 90\% confidence level as a function of $m_{A^\prime}$ using a Bayesian method~\cite{Zhu:2008ca}. The multiplicative systematic uncertainty is incorporated by smearing the likelihood curve using a Gaussian function with a width representing the systematic uncertainty as follows:
\begin{equation}
    L'(\mathcal{B})=\int_{0}^{1}L\left(\frac{\mathcal{\epsilon}}{\hat{\epsilon}}\mathcal{B} \right)\exp \left(-\frac{\left[\epsilon-\hat{\epsilon}\right]^2}{2\sigma^{2}_{s}} \right)d\epsilon, 
\end{equation}
where $L$ and $L^\prime$ are the likelihood curves before and
after taking into account the systematic uncertainty; $\epsilon$, $\hat{\epsilon}$ and  $\sigma_{S}$ are the detection efficiency, the nominal efficiency, and the absolute total systematic uncertainty on the efficiency, respectively. 
As shown in Fig.~\ref{aeta}, the combined limits on the production 
rate $\mathcal{B}(\eta\rightarrow\gamma \mathbf{\it A'})\times\mathcal{B}(\mathbf{\it A'}\rightarrow e^{+}e^{-})$ are established at the level 
of $(1-70)\times 10^{-6}$ for $0 <m_{A^\prime}<0.4$ GeV/c$^2$ in $\eta\rightarrow\gamma e^+e^-$, and 
$\mathcal{B}(\eta'\rightarrow\gamma \mathbf{\it A'})\times\mathcal{B}(\mathbf{\it A'}\rightarrow e^{+}e^{-})$  are constrained at the level of 
$(0.5-3.5)\times 10^{-6}$ for $0 <m_{A^\prime}<0.7$ GeV/c$^2$ in $\eta'\rightarrow\gamma e^+e^-$. 
With the expected dark photon decay branching fraction of $A'\rightarrow e^{+}e^{-}$ obtained from Ref.~\cite{Batell:2009yf}, the upper limits on the coupling  strength $\varepsilon$ both vary in the range of $10^{-2}-10^{-3}$.

\begin{figure*}[htbp]
    \centering
    \subfigure{
        \includegraphics[scale=0.42]{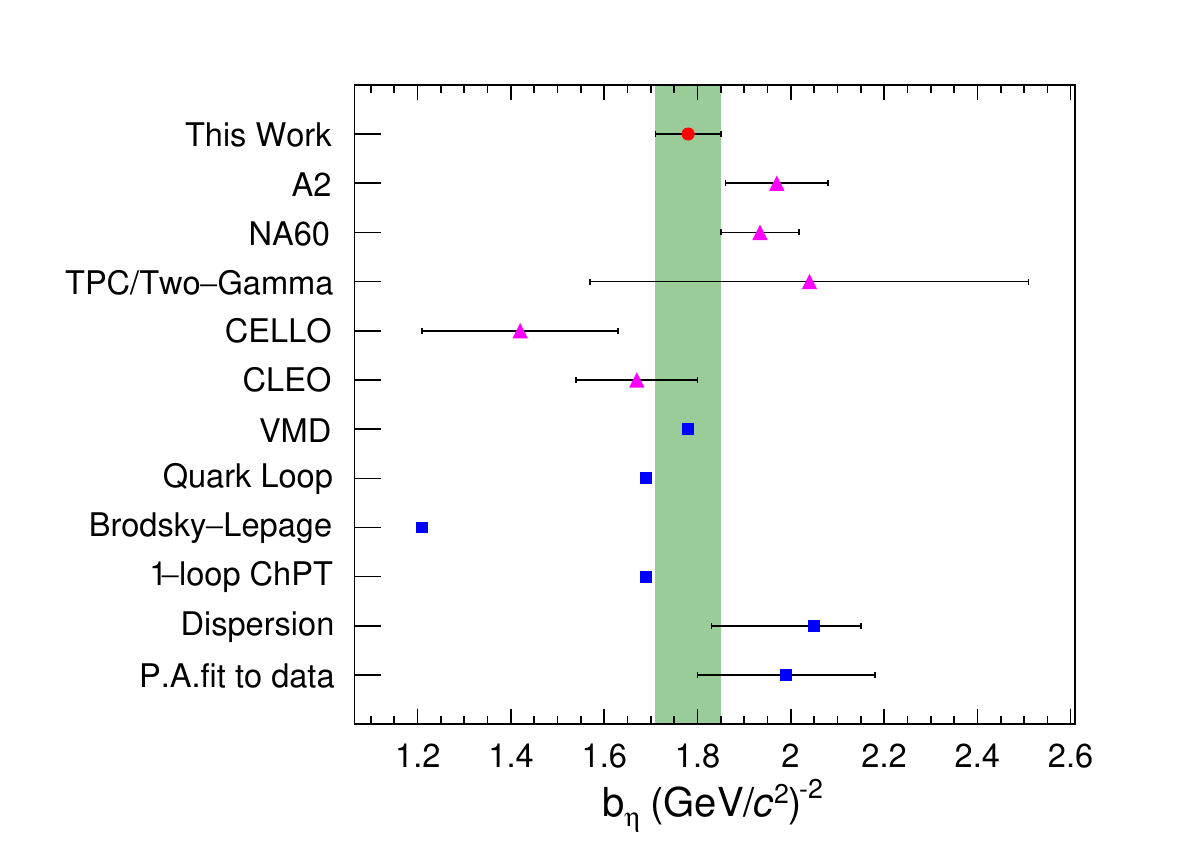}
    }
    \subfigure{
        \includegraphics[scale=0.42]{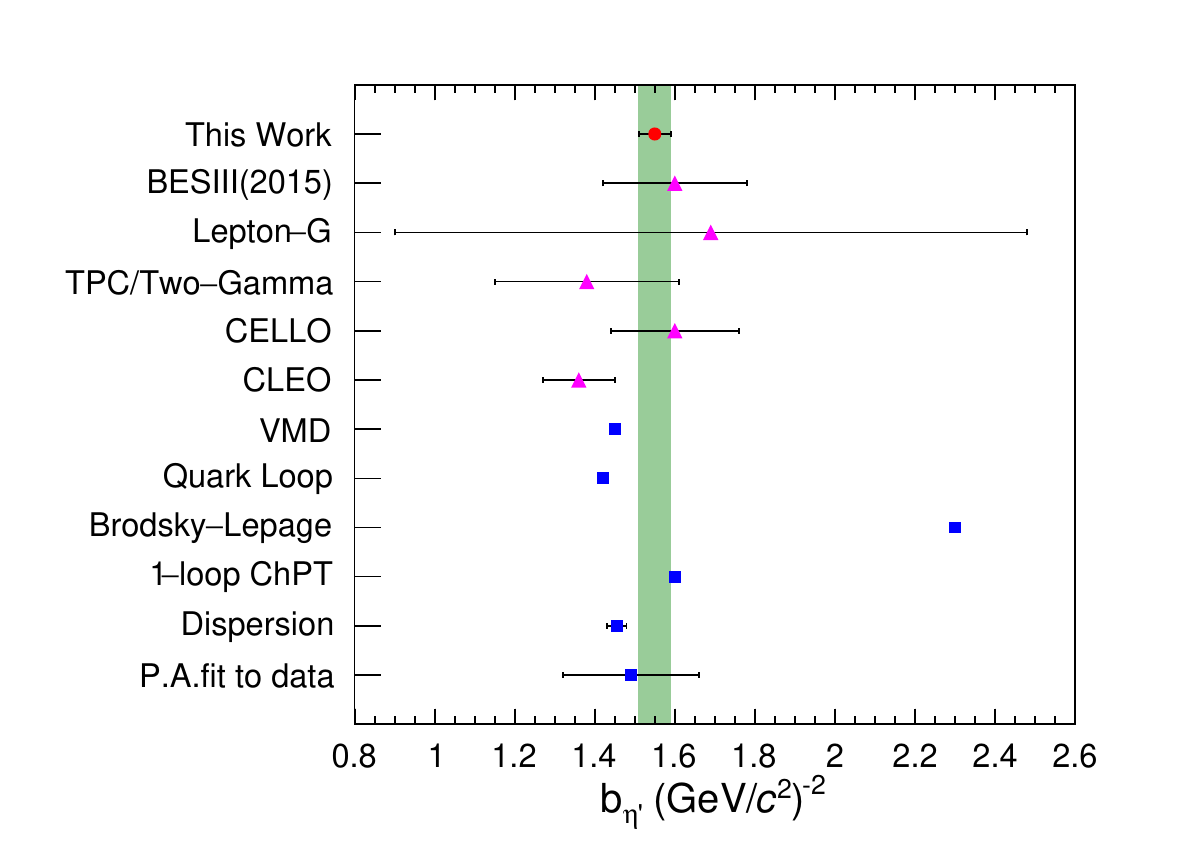}
    }
    \caption{Slope parameters of the $\eta$(left panel) and $\eta^\prime$(right panel) TFFs extracted from different experiments and calculated by different theoretical models. The points refer to experiments~\cite{Adlarson:2016hpp, NA60:2016nad, TPCTwoGamma:1990dho, CELLO:1990klc, CLEO:1997fho, Dzhelyadin:1979za, BESIII:2015zpz} (purple triangles), theoretical calculations~\cite{Landsberg:1985gaz, Hanhart:2013vba, Ametller:1991jv, Escribano:2013kba, Holz:2022hwz, Ametller:1983ec} (blue squares), and this work (red dots). The green bands are the total uncertainties of this work.}
    \label{comtff}
\end{figure*}

\section{summary}
Using a sample of 10 billion $J/\psi$ events collected by the BESIII detector, the Dalitz decays $\eta\rightarrow\gamma e^{+}e^{-}$ and $\eta^\prime\rightarrow\gamma e^{+}e^{-}$ are studied using $J/\psi$ radiative decays.
The branching fractions are determined to be $\mathcal{B}(\eta\rightarrow\gamma e^{+}e^{-}) = (7.07 \pm 0.05 \pm 0.23)\times10^{-3}$ and $\mathcal{B}(\eta'\rightarrow\gamma e^+e^-) = (4.83\pm0.07\pm0.14)\times10^{-4}$, respectively.
%which are both consistent with theoretical calculations~\cite{Landsberg:1985gaz} and previous measurements~\cite{CELSIUSWASA:2007ifz,BESIII:2015zpz}. 
 
We also measure the TFF as a function of $M(e^{+}e^{-})$ with a single-pole parametrization for the $\eta$ and a multipole parametrization for the $\eta'$. The parameters $\Lambda_{\eta}$, $\Lambda_{\eta'}$, and $\gamma_{\eta'}$ are determined to be $(0.749 \pm 0.027 \pm 0.008)~ {\rm GeV}/c^{2}$, $(0.802 \pm 0.007\pm 0.008)~ {\rm GeV}/c^{2}$ and $(0.113\pm0.010\pm0.002)~ {\rm GeV}/c^{2}$, respectively, in good agreement with previous works, as shown in the Fig~\ref{comtff}. The corresponding radii of interaction region of the $\eta$ and $\eta'$ mesons are calculated to be $R_{\eta}=(0.645 \pm 0.023 \pm 0.007 )~ {\rm fm}$ and $R_{\eta'}=(0.596 \pm 0.005 \pm 0.006)~ {\rm fm}$. 
 
In addition, we search for a dark photon $A^\prime$ in $\eta\rightarrow \gamma e^+e^-$ and $\eta^\prime\rightarrow \gamma e^+e^-$, and the significance for each case is less than $0.5\sigma$.
The upper limits on their branching fractions, as illustrated in Fig.~\ref{aeta}, are calculated to be $(1-70) \times 10^{-6}$ at the 90\% confidence level in the range of $0<m_{A^\prime}<0.4$ GeV/$c^2$ and  $(0.5-3.5) \times 10^{-6}$ at the 90\% confidence level in the range of $0<m_{A^\prime}<0.7$ GeV/$c^2$, respectively. The upper limits of the coupling strength $\varepsilon$ vary in the range of $10^{-2}-10^{-3}$.

\section*{ACKNOWLEDGMENTS}
The BESIII Collaboration thanks the staff of BEPCII and the IHEP computing center for their strong support. This work is supported in part by National Key R\&D Program of China under Contracts Nos. 2020YFA0406300, 2020YFA0406400; National Natural Science Foundation of China (NSFC) under Contracts Nos. 11635010, 11735014, 11835012, 11935015, 11935016, 11935018, 11961141012, 12005195, 12025502, 12035009, 12035013, 12061131003, 12192260, 12192261, 12192262, 12192263, 12192264, 12192265, 12221005, 12225509, 12235017; the Chinese Academy of Sciences (CAS) Large-Scale Scientific Facility Program; the CAS Center for Excellence in Particle Physics (CCEPP); Joint Large-Scale Scientific Facility Funds of the NSFC and CAS under Contract No. U1832207; CAS Key Research Program of Frontier Sciences under Contracts Nos. QYZDJ-SSW-SLH003, QYZDJ-SSW-SLH040; 100 Talents Program of CAS; The Institute of Nuclear and Particle Physics (INPAC) and Shanghai Key Laboratory for Particle Physics and Cosmology; European Union's Horizon 2020 research and innovation programme under Marie Sklodowska-Curie grant agreement under Contract No. 894790; German Research Foundation DFG under Contracts Nos. 455635585, Collaborative Research Center CRC 1044, FOR5327, GRK 2149; Istituto Nazionale di Fisica Nucleare, Italy; Ministry of Development of Turkey under Contract No. DPT2006K-120470; National Research Foundation of Korea under Contract No. NRF-2022R1A2C1092335; National Science and Technology fund of Mongolia; National Science Research and Innovation Fund (NSRF) via the Program Management Unit for Human Resources \& Institutional Development, Research and Innovation of Thailand under Contract No. B16F640076; Polish National Science Centre under Contract No. 2019/35/O/ST2/02907; The Swedish Research Council; U. S. Department of Energy under Contract No. DE-FG02-05ER41374

\bibliographystyle{apsrev4-2}
\bibliography{main}

\end{document}